\documentclass[12pt,a4paper]{article}

\usepackage{epsfig}
\usepackage{amsmath,amsfonts,amssymb}
\usepackage{dsfont}
\usepackage{t1enc}
\usepackage{cite}
\usepackage{colortbl}
\usepackage{pifont}
\usepackage{slashed}

\providecommand{\openone}{\leavevmode\hbox{\small1\kern-3.8pt\normalsize1}}

\usepackage{colortbl,colordvi,xcolor,color}
\usepackage{rotating}

\usepackage[T1]{fontenc} 

\newcommand{\refeq}[1]{(\ref{#1})}

\numberwithin{equation}{section}
\newcommand{\nn}{\nonumber \\}
\newcommand{\beq}{\begin{equation}}
\newcommand{\eeq}{\end{equation}}
\newcommand{\ba}{\begin{array}}
\newcommand{\ea}{\end{array}}
\newcommand{\bea}{\begin{eqnarray}}
\newcommand{\eea}{\end{eqnarray} }
\newcommand{\bal}{\begin{align}}
\newcommand{\eal}{\end{align}}

\newcommand{\tr}{\mathrm{tr}}
\newcommand{\ga}{\gamma}

\newcommand{\kslash}{\slashed{k}}
\newcommand{\pslash}{\slashed{p}}

\renewcommand{\d}{\partial}

\newcommand{\eps}{\epsilon}

\begin{document}

\begin{flushright}
{CERN-TH-2018-058~~~~~~~~}
\end{flushright}

\vspace*{1cm}

\begin{center}
\begin{Large}
{\bf Dimensional regularization vs methods in fixed dimension with and without $\gamma_5$}
\end{Large}

\vspace{0.5cm}
A.\ M.\ Bruque$^{a}$, A.\ L.\ Cherchiglia$^{b}$  and M. P\'erez-Victoria$^{a,c}$  \\[0.2cm] 
\end{center}
{ \it  
${}^a$  CAFPE and Departamento de F\'{\i}sica Te\'orica y del Cosmos, 
Universidad de Granada, Campus de Fuentenueva, E-18071 Granada, Spain  \\
${}^b$ Centro de Ci\^ecias Naturais e Humanas, Universidade Federal do ABC, Santo Andr\'e, Brazil \\
${}^c$ Theoretical Physics Department, CERN, Geneva, Switzerland 
}

\begin{abstract}
We study the Lorentz and Dirac algebra, including antisymmetric $\epsilon$ tensors and the $\gamma_5$ matrix, in implicit gauge-invariant regularization/renormalization methods defined in fixed integer dimensions. They include constrained differential, implicit and  four-dimensional renormalization. We find that these fixed-dimension methods face the same difficulties as the different versions of dimensional regularization. We propose a consistent procedure in these methods, similar to the consistent version of regularization by dimensional reduction. 
\end{abstract}

\section{Introduction}
\label{sec:intro} 

It is well known that regularization methods based on analytical continuation in a complex dimension $d$ face some problems in the presence of $\gamma_5$ matrices and completely antisymmetric $\epsilon$ tensors. The reason is that the usual properties of these objects in integer dimension $n$ are not consistent with the treatment of Lorentz tensors in dimensional regularization (DReg)~\cite{DReg,tHV}.  Therefore, one has to give up some of these properties~\cite{BM}. In particular, a consistently-defined $\gamma_5$ that approaches the standard $\gamma_5$ as $d\to n$ cannot anticommute with the Dirac matrices in DReg~\cite{BM,Bonneau} \footnote{An anticommuting $\gamma_5$ is often employed in DReg, nevertheless. Although this may be safe for many calculations, as shown in~\cite{Jegerlehner:2000dz}, a well-defined renormalized theory requires a consistent unified treatment of arbitrary diagrams and amplitudes. In particular, this is necessary to prove important properties of the theory to all orders.} and there is no finite-dimensional complete set in Dirac space---which is an obstacle, for instance, for Fierz reorderings and supersymmetry. These complications are related to the fact that Lorentz covariants in complex dimension $d$ are treated as formal objects, in which the indices do not take actual values. Even if quantities such as  $\{\gamma_5,\gamma_\mu\}$ are evanescent, i.e.\ they approach zero as $d\to n$, due to the poles at $d=n$ in the loop integrals they leave a vestige in the renormalized functions after minimal subtraction (MS).

A consistent set of rules in DReg to manipulate Lorentz tensors, including the $\epsilon$ tensor and $\gamma_5$ matrix, was proposed by Breitenlohner and Maison in~\cite{BM}. They used the original definition of $\gamma_5$ by t'Hooft and Veltman (tHV)~\cite{tHV}. Completed with these rules and MS (or $\overline{\text{MS}}$), DReg provides a consistent renormalization scheme. However, besides genuine anomalies, spurious anomalies appear in some correlators of axial vector currents in QCD and chiral gauge theories, including the Standard Model~\cite{ChanowitzFurmanHinchliffe}. These pose no fundamental problem, as it has been shown that they can be eliminated by an additional finite subtraction in a systematic way~\cite{CostaTonin}. But such a correction represents a complication in explicit calculations. This is the main reason for looking for alternatives to the tHV definition of $\gamma_5$. Furthermore, DReg explicitly breaks supersymmetry, so it is not a convenient method in supersymmetric theories. 

An alternative set of rules for Lorentz tensors and Dirac matrices, which define regularization by dimensional reduction (DRed), was proposed by Siegel in~\cite{Siegel1} with the purpose of preserving supersymmetry. In this case, the $\eps$ tensor, the Dirac gamma matrices and the $\gamma_5$ matrix are the original $n$-dimensional objects and thus the Dirac algebra is in principle performed in integer dimensions. The name of the method indicates that when these objects are contracted with tensors associated to the dimensionally-regularized integrals, they are projected into the formal $d$-dimensional Minkowski space. But due to this projection, the conflict between the $n$-dimensional relations and the $d$-dimensional Lorentz space reappears. In fact, Siegel himself showed in~\cite{Siegel2} that the set of rules in the original formulation is inconsistent. A consistent set of rules can be defined by, once again, giving up genuine $n$-dimensional relations that require giving explicit integer values to the Lorentz indices~\cite{Avdeev,Avdeev2,Stoeckinger}. Unsurprisingly, this consistent version of DRed does not preserve supersymmetry. The four-dimensional helicity method (FDH) is a variant of DRed that treats external vector fields as strictly four dimensional. For our purposes we need not distinguish it from DRed. 

In view of the unavoidable difficulties of the dimensional methods when applied to chiral theories or chiral operators, it is reasonable to turn to methods defined in the fixed dimension of interest (often $n=4$). Indeed, none of the issues discussed above seems at first sight to be relevant when the original dimension is kept fixed. However, in this paper we show that this expectation is too na\"ive. It turns out that the formal treatment of Lorentz tensors and Dirac space in certain efficient fixed-dimension methods shares many properties with the one in dimensional methods. As a consequence, the same consistency problems with $\epsilon$ and $\gamma_5$ are found. Consistency can be recovered, once again, by giving up some $n$-dimensional relations. The resulting methods do not preserve supersymmetry.

Of course, in explicit regularizations in fixed dimension $n$, such as a momentum ultraviolet cutoff or those based on a modification of the Lagrangian, the $n$-dimensional Lorentz and Dirac manipulations inside or outside loop integrals are well defined. The same is true at the renormalized level in physical renormalization schemes defined by renormalization conditions. However, when some {\it a priori\/} subtraction prescription is used (similar to MS in DReg), the commutation of the usage of $n$-dimensional identities with the substraction of divergences needs to be checked. This is specially so in fixed-dimension methods that directly provide renormalized amplitudes without explicit counterterms. Here we are interested in methods of this kind with the potential of satisfying the quantum action principle~\cite{QAP}, from which basic properties, such as Ward identities, follow to all orders. We will generically refer to them as {\em implicit methods}. Specifically, we study in detail three similar implicit methods: constrained differential regularization/renormalization (CDR)~\cite{CDR}, constrained implicit regularization/renormalization (CIReg)~\cite{IR} and four-dimensional regularization/renormalization~(FDR)~\cite{FDR}.  These methods have been applied to one-loop and multi-loop calculations in QCD~\cite{PerezVictoria:1998fj,Sampaio:2005pc,subintegrationFDR}, the Standard Model~\cite{Cherchiglia:2012zp,Donati:2013iya,Pittau:2013qla}, supersymmetric models~\cite{Guasch:2001kz,Mas:2002xh,Carneiro:2003id} and supergravity~\cite{delAguila:1997yd}, among other theories. In all these examples, gauge invariance and supersymmetry, when relevant, have been preserved. We will also make some comments about the four-dimensional formalism (FDF)~\cite{FDF} of FDH, which shares some features with FDR.

The first of these methods was originally defined in position space, as a gauge-invariant version of differential renormalization~\cite{DiffRen}, but it works equally well in momentum space. Actually, its momentum-space version is implemented in {\tt FormCalc}~\cite{FormCalc}.\footnote{{\tt FormCalc} has the option of working in $d$ or 4 dimensions, that is, in DReg or CDR. The latter is more suitable for supersymmetric theories.} CDR gives renormalized expressions without any intermediate regularization, essentially by an implicit subtraction of local singularities (polynomial in external momenta, in agreement with Weinberg's theorem). CIReg and FDR work in momentum space at the integrand level. Both methods use straightforward partial-fraction identities to isolate the ultraviolet divergences, with no external momenta in the denominators. The divergent parts are then ignored, that is, subtracted. Again, no regularization is necessary. One difference between them is that CIReg keeps physical masses in the divergent parts, while FDR does not, but these include an auxiliary mass $\mu$, which is introduced before the algebraic manipulations to avoid artificial infrared divergences and taken to zero at the end.\footnote{CIReg can also be implemented without masses in divergent parts \cite{forestCIR}. In this mass-independent scheme, and in all cases in massless theories, the scale $\mu$ is introduced in CIReg as well, but only in denominators.} This scale is essential in FDR and will be very relevant in the discussion below. Let us stress that, notwithstanding its name, FDR can be used in the very same manner in any integer dimension. All three methods can in principle deal as well with genuine infrared divergences, but only FDR has been studied in detail in this context, both for virtual and real singularities~\cite{FDR,Donati:2013iya}. The equivalences in non-chiral theories and at the one-loop level of CDR, CIReg (in a massless scheme) and DRed have been established in~\cite{FormCalc,delAguila:1999bd} and~\cite{Pontes:2007fg}. Concerning the preservation of unitarity and locality in multiloop calculations without counterterms, CDR and CIReg rely on Bogoliubov's recursive renormalization and Zimmermann's forest formula~\cite{Bogoliubov:1957gp,Hepp:1966eg,forest,forestCDR,forestCIR}. In the literature of FDR, {\em sub-integration consistency\/}  is checked for different structures and imposed by an extra finite subtraction of sub-diagrams~\cite{subintegrationFDR}. A systematic implementation of this idea should eventually be equivalent to the application of the forest formula. At any rate, here we are concerned with the treatment of Lorentz tensors and Dirac matrices in these fixed-dimension methods, and one-loop examples will be sufficient to illustrate our main messages.

The paper is organized as follows. In section~\ref{s:Lorentz}, we describe the treatment of Lorentz tensors in DReg/DRed and in implicit methods. We stress the fact that, in order to preserve basic properties of the integrals, the contraction of Lorentz indices cannot commute with renormalization. We  also explain how this requirement is implemented in the different methods.  In section~\ref{s:inconsistent} we show a consequence of it: some identities that are valid in standard $n$-dimensional spaces are spoiled by the renormalization process. Thus, using these identities lead to inconsistent results.  
In section~\ref{s:Dirac} we study how the Dirac algebra is affected by these potential inconsistencies. We find that implicit methods have problems with the Dirac algebra in odd dimensions and with the $\gamma_5$ matrix in even dimensions. These issues parallel the ones in dimensional methods. In section~\ref{s:formal} we propose a well-defined procedure that avoids inconsistencies in implicit methods. This procedure is analogous to the consistent version of DRed. We discuss allowed simplifications within this scheme, including shortcuts that have already been used in FDR. We give simple examples in $n=2$ and $n=4$ in section~\ref{s:examples} and conclude in section~\ref{s:conclusions}. An appendix collects functions that appear in our explicit calculations. In order to keep the equations as short and simple as possible, formulas are often given in $n=2$.  We work in Euclidean space and formal generalizations of it.

\section{Lorentz tensors and index contraction}
\label{s:Lorentz}
In dimensional methods, the contraction of Lorentz indices in a tensorial integral does not, in general, commute with regularization and renormalization. This comes from the simple fact that the trace of the $d$-dimensional metric tensor is $\delta_{\mu \mu}=d=n-\varepsilon \neq n$. When it hits a pole $1/\varepsilon$ in a divergent integral, the term linear in $\varepsilon$ will give rise to a finite contribution, which is not subtracted in MS and survives when $\varepsilon$ is taken to zero.

We show next that, actually, index contraction does not commute with renormalization in any gauge-invariant method that consistently replaces each overall-divergent integral by a unique finite expression. CDR, CIReg or FDR belong to this class. The proof of the quantum action principle in perturbation theory relies on two non-trivial properties: invariance under shifts of the integration momenta and numerator-denominator consistency. The first property is related to translational invariance and guaranties independence of momentum rooting. The second one requires that the application of the kinetic operator to the propagator associated to some line in a Feynman graph is equivalent to pinching of that line, that is, its contraction to a point. This is necessary for a consistent treatment of the quadratic and interaction terms in perturbation theory~\cite{BM}. These properties need not hold in arbitrary definitions of regularized or subtracted integrals.

Shift invariance can be related to the vanishing of total derivatives with respect to integration momenta: 
\beq
0 = \left[ \int d^nk \,\left(f(k+p) - f(k)\right)\right]^R = p_\nu \left[ \int d^nk \, \frac{\d} {\d k_\nu} f(k) \right]^R + O(p^2) . \label{shift}
\eeq
Here, $R$ indicates that the expression inside the corresponding brackets is renormalized, i.e.\ subtracted and with any possible regulator or auxiliary parameter removed (except for the unavoidable renormalization scale). We require that the operation $[.]^R$ be linear:
\beq
\left[aF+bG\right]^R= a \left[F\right]^R + b \left[G\right]^R, 
\eeq
where $a,b$ are numbers or external objects, such as external momenta. This holds in all the methods we study in this paper. Consider the following two-dimensional integral:
\begin{align}
f_{\mu\nu} &= \int d^2k\, \frac{\d}{\d k_\mu} \frac{k_\nu}{k^2+m^2} \nn
&= \int d^2 k \, \left(\frac{\delta_{\mu\nu}}{k^2+m^2} - 2 \frac{k_\mu k_\nu}{(k^2+m^2)^2} \right) \label{fmunu}
\end{align}
According to~\refeq{shift}, shift invariance requires $[f_{\mu\nu}]^R=0$, and thus, calling
\beq
I_{\mu\nu} = \int d^2 k \, \frac{k_\mu k_\nu}{(k^2+m^2)^2} ,
\eeq
we have 
\begin{align}
\left[ I_{\mu\nu} \right]^R & = \frac{1}{2} \delta_{\mu\nu} \left[ \int d^2 k \, \frac{1}{k^2+m^2}  \right]^R \nn
& = \frac{1}{2} \delta_{\mu\nu} \left( \left[\int d^2k \, \frac{k^2}{(k^2+m^2)^2} \right]^R +  \left[\int d^2k \, \frac{m^2}{(k^2+m^2)^2} \right]^R  \right) \nn
& = \frac{1}{2} \delta_{\mu\nu} \left( \left[ I_{\alpha\alpha} \right]^R+\pi \right) .  \label{Imunu}
\end{align}
That is, shift invariance forbids symmetric integration (in $n$ dimensions). 
In the second line we have used numerator-denominator consistency, $(k^2+m^2)/(k^2+m^2)=1$. This looks trivial in the formal equations above, but it is not so in methods that modify the propagators at intermediate steps of the calculation. In the third line we have assumed that integrals finite by power counting are not changed by renormalization. This assumption is essential in the definition of dimensional regularization and also in the definition of CDR, CIReg and FDR, as should already be clear from the brief explanations in the introduction. We can rewrite~\refeq{Imunu} as
\beq
\delta_{\mu\nu} \left[ I_{\mu\nu} \right]^R = \left[ \delta_{\mu\nu} I_{\mu\nu} \right]^R + \pi \label{notrace}.
\eeq
So, we see that {\em renormalization does not commute with index contraction if it commutes with shifts of integration momenta and respects numerator-denominator consistency.} This is in fact the origin of trace anomalies~\cite{traceanomallies} and also of chiral anomalies, as we shall see. The same conclusion can be proven in arbitrary integer dimension $n$ using similar arguments. 

Let us now examine how the different renormalization methods we are discussing recover \refeq{Imunu}, and thus comply with~\refeq{shift}. In the case of dimensional methods, we have
\begin{align}
\left[I_{\mu\nu}\right]^R & =  \left[ \int d^d k \, \frac{k_\mu k_\nu}{(k^2+m^2)^2} \right]^{S} \nn
& = \left [\int d^d k \,  \frac{1}{d} \delta_{\mu\nu} \frac{ k^2}{(k^2+m^2)^2}\right]^S \nn
& = \left[ \int d^d k \, \left( \frac{1}{2} + \frac{\varepsilon}{4} + O(\varepsilon^2) \right) \delta_{\mu\nu} \frac{ k^2}{(k^2+m^2)^2} \right]^S \nn
& = \left[\frac{1}{2} \delta_{\mu\nu} \int d^d k \, \frac{k^2}{(k^2+m^2)^2}  + \left(\frac{\varepsilon}{4}+O(\varepsilon^2)\right) \delta_{\mu\nu}  \left(2 \pi \frac{1}{\varepsilon} + O(\varepsilon^0) \right) \right]^S \nn
& = \frac{1}{2} \delta_{\mu\nu} \left( \left[ I_{\alpha\alpha}\right]^R + \pi \right),
\end{align}
in agreement with~\refeq{Imunu}. Here, $S$ indicates MS followed by $\varepsilon \to 0$. Note that before the $S$ operation, $\delta$ is the Euclidean metric in $d$ formal dimensions, which satisfies $\delta_{\mu\mu}=d$.


In CDR, the finite local terms in the renormalized value of the different overall-divergent tensor integrals are fixed by requiring compatibility with shift invariance and numerator-denominator consistency. Hence, $[f_{\mu\nu}]^R=0$ by construction and the extra local term in the tensor integral is fixed just as in equation~\refeq{notrace}. 

CIReg has the advantage of working at the integrand level. Tensor integrands are expressed as simpler integrands plus total derivatives. Integrating the latter gives potential surface terms, which are dropped by definition. So, shift invariance is enforced by the very definition of the method. For instance, using the same relation as in~\refeq{fmunu},
\begin{align}
\left[I_{\mu\nu}\right]^R &=  \left[\int d^2 k \, \left(\frac{1}{2}\frac{\delta_{\mu\nu}}{k^2+m^2} -  \frac{1}{2}\frac{\d}{\d k_\mu} \frac{k_\nu}{k^2+m^2}\right) \right]^R \nn
& =\frac{1}{2}{\delta_{\mu\nu}} \left[\int d^2 k \, \frac{1}{k^2+m^2} \right]^R \nn
& = \frac{1}{2} \delta_{\mu\nu} \left( \left[ I_{\alpha\alpha}\right]^R + \pi \right)
. \label{tensorCIR}
\end{align}
We see that the same local terms as in CDR are found, but in this case there is a simple prescription to obtain them. 
Obviously $[f_{\mu\nu}]^R=0$ and~\refeq{notrace} is satisfied. At this point, it is important to make the following observation. We can also write
\beq
\left[I_{\alpha\alpha} \right]^R = \left[\int d^2 k \, \left(\frac{1}{k^2+m^2} -  \frac{1}{2}\frac{\d}{\d k_\alpha} \frac{k_\alpha}{k^2+m^2}\right) \right]^R . \label{contractedST}
\eeq
Dropping the second term would contradict~\refeq{tensorCIR}. Accordingly, CIReg does not drop this sort of surface term when the index in the total derivative is contracted with a loop momentum. Therefore, just as CDR, CIReg distinguishes by definition contracted and non-contracted Lorentz indices. Note that the vanishing of the second term on the right hand of~\refeq{contractedST} is not necessary for shift invariance: in~\refeq{shift} the index in the total derivative is contracted with the index in the (external) momentum shift, so it can never be contracted with the index of a loop momentum. 

In FDR, which also works at the integrand level, the extra local terms necessary for shift invariance result automatically from the introduction of the scale $\mu$, together with some additional prescriptions. In this method, 
\begin{align}
\left[ I_{\mu\nu}  \right]^R & = \left[ \int d^2k \, \frac{k_\mu k_\nu}{(k^2+\mu^2+m^2)^2}  \right]^S \nn
& = \frac{1}{2} \delta_{\mu\nu}  \left[ \int d^2 k\, \frac{k^2}{(k^2+\mu^2+m^2)^2} \right]^S \nn
& =  \frac{1}{2} \delta_{\mu\nu}  \left[ \int d^2 k\, \frac{k^2+\mu^2}{(k^2+\mu^2+m^2)^2} - \int d^2 k\, \frac{\mu^2}{(k^2+\mu^2+m^2)^2}\right]^S \nn
& = \frac{1}{2} \delta_{\mu\nu} \left(\left[ I_{\alpha\alpha} \right]^R -  \left[ \int d^2 k \, \frac{\mu^2}{(k^2+ \mu^2+m^2)^2} \right]^S  \right)\nn
& = \frac{1}{2} \delta_{\mu\nu} \left(\left[ I_{\alpha\alpha} \right]^R +  \pi  \right). \label{tensorFDR}
\end{align}
Several explanations are in order. The first step in FDR is the introduction of the scale $\mu$, as done in the first line of~\refeq{tensorFDR}. The symbol $[.]^S$ in this case refers to the FDR subtractions, followed by the limit $\mu \to 0$ (outside logarithms). In the second line, we have used the property of symmetric integration, which is allowed in this method {\em after} the scale $\mu$ has been introduced. In the forth line we have used the so-called {\em global prescription\/} of FDR, according to which the possible $k^2$ in numerators inside $[.]^R$ should be also replaced by $k^2+\mu^2$, just as in the denominators. As emphasized in~\cite{FDR}, this is necessary to preserve numerator-denominator consistency. Finally, the integral in the second term of the fourth line of~\refeq{tensorFDR} is finite and goes to zero as $\mu\to 0$. However, a nonvanishing contribution is found as shown in the last line, because FDR performs an oversubtraction, treating this integral as divergent (for power counting, $\mu$ is counted like an integration momentum). In the FDR language integrals of this kind are called {\em extra integrals}. They play the same role as the extra local terms in CDR, with the advantage that the necessary terms arise directly from a simple and universal prescription, formulated without reference to specific integrals.  The result in~\refeq{tensorFDR} coincides with the one in the previous methods, as it should to guarantee $[f_{\mu\nu}]^R=0$, and thereby shift invariance. 

Let us summarize this section. Just as in dimensional renormalization, the contraction of Lorentz indices does not commute with renormalization in the implicit methods we are considering, which respect invariance under shifts of the integration momenta and numerator-denominator consistency. In the latter methods, $k^2$ and $k_\mu k_\nu$ have to be treated in a different manner by hand. This requires writing the diagrams in some normal form that allows for a unique identification of tensors with contracted and uncontracted indices. 

\section{Relations in genuine integer dimension}
\label{s:inconsistent}
Genuine $n$-dimensional identities (GnDI) spoil the uniqueness of the normal form and thus can lead to inconsistencies in implicit methods, which parallel the ones in DRed.  By GnDI we mean equalities depending crucially on the fact that the Lorentz indices can take $n$ different integer values. 
Consider the determinant
\beq
\mathrm{Det}(\mu_1\dots \mu_m;\nu_1 \dots \nu_m) \equiv 
\left| \begin{array}{cccc}
\delta_{\mu_1\nu_1} & \delta_{\mu_1\nu_2}  & \dots & \delta_{\mu_1 \nu_m} \\
\delta_{\mu_2\nu_1} & \delta_{\mu_2 \nu_2} & \dots & \delta_{\mu_2 \nu_m} \\
\vdots & \vdots & & \vdots \\
\delta_{\mu_m \nu_1} & \delta_{\mu_m\nu_2} & \dots & \delta_{\mu_m\nu_m}
\end{array}   \right| \, . \label{determinant}
\eeq
In standard algebra, this object vanishes when $m>n$, since it is then unavoidable to have at least two identical rows, as the indices can take only $n$ different values. However, this is not necessarily true when used inside $[.]^R$, because contracted and uncontracted indices are treated differently if index contraction does not commute with renormalization. To show this more explicitly, let us consider the case with $n=2$ and $m=3$. Requiring the determinant~\refeq{determinant} to vanish we  have
\begin{align}
0 & = \left[  0 \right]^R \nn
& \overset{?}{=} \left[ \text{Det}(\alpha\mu\nu;\beta\rho\sigma) p_{1\mu} p_{2\nu} p_{3\rho} p_{4\sigma}  I_{\alpha\beta} \right]^R \nn
& = (p_1\cdot p_3 \, p_2 \cdot p_4 - p_1\cdot p_4 \, p_2 \cdot p_3) \left[I_{\alpha\alpha} \right]^R  -  p_{1\mu}p_{3\rho} \, p_2\cdot p_4  \left[I_{\rho\mu}\right]^R \nn
& \phantom{=} \mbox{}  +  p_{3\rho}p_{2\nu} \, p_1\cdot p_4  \left[I_{\rho\nu}\right]^R+ p_{4\sigma}p_{1\mu} \, p_2\cdot p_3  \left[I_{\sigma\mu} \right]^R-  p_{2\nu}p_{4\sigma}\,  p_1\cdot p_3 \left[I_{\sigma\nu}\right]^R.   \label{nonzerodet}
\end{align}
If we now use~\refeq{Imunu}, we find
\beq
0 \overset{?}{=} \pi (p_1\cdot p_4 \, p_2 \cdot p_3 - p_1\cdot p_3 \, p_2 \cdot p_4) ,  \label{bad}
\eeq 
which is obviously not true for general $p_i$. 

This simple example is sufficient to prove the main assertion of this paper: {\em Using GnDI before renormalization can lead to inconsistencies in implicit methods.}  The origin of this issue is the non-commutation of index contraction with renormalization. The difficulties with $\gamma_5$, discussed in the next section, are a direct consequence of it.

In dimensional methods, it is clear that the determinant~\refeq{determinant} does not vanish if $\delta$ is the $d$-dimensional metric, so obviously the second equality in~\refeq{nonzerodet} is invalid. However, an $n$-dimensional metric $\bar{\delta}$ ($\tilde{\delta}$) is introduced in DReg (DRed), with $\bar{\delta}_{\mu\mu}=\tilde{\delta}_{\mu\mu}=n$. The relation between the $n$-dimensional and $d$-dimensional metrics is different in DReg and DRed:
\begin{align}
& \delta_{\mu\nu} \bar{\delta}_{\nu\rho} = \bar{\delta}_{\mu\rho}, ~~~~~\text{(DReg)}\label{projectionDReg} ; \\
& \delta_{\mu\nu}  \tilde{\delta}_{\nu\rho} = {\delta}_{\mu\rho}, ~~~~~\text{(DRed)}\label{projection} .
\end{align}
Let us define
\beq
\mathrm{\overline{Det}}(\mu_1\dots \mu_m;\nu_1 \dots \nu_m) \equiv 
\left| \begin{array}{cccc}
\bar{\delta}_{\mu_1\nu_1} & \bar{\delta}_{\mu_1\nu_2}  & \dots & \bar{\delta}_{\mu_1 \nu_m} \\
\bar{\delta}_{\mu_2\nu_1} & \bar{\delta}_{\mu_2 \nu_2} & \dots & \bar{\delta}_{\mu_2 \nu_m} \\
\vdots & \vdots & & \vdots \\
\bar{\delta}_{\mu_m \nu_1} & \bar{\delta}_{\mu_m\nu_2} & \dots & \bar{\delta}_{\mu_m\nu_m}
\end{array}   \right| \, . \label{determinantbar}
\eeq
For $n=2$, in DReg we have
\begin{align}
0 & = \left[  0 \right]^R \nn
& = \left[ \overline{\text{Det}}(\alpha\mu\nu;\beta\rho\sigma) p_{1\mu} p_{2\nu} p_{3\rho} p_{4\sigma}  I_{\alpha\beta} \right]^R \nn
& = (p_1\cdot p_3 \, p_2 \cdot p_4 - p_1\cdot p_4 \, p_2 \cdot p_3) \left[\bar{\delta}_{\alpha\beta}I_{\alpha\beta} \right]^R  -  p_{1\mu}p_{3\rho} \, p_2\cdot p_4  \left[I_{\rho\mu}\right]^R \nn
& \phantom{=} \mbox{}  +  p_{3\rho}p_{2\nu} \, p_1\cdot p_4  \left[I_{\rho\nu}\right]^R+ p_{4\sigma}p_{1\mu} \, p_2\cdot p_3  \left[I_{\sigma\mu} \right]^R-  p_{2\nu}p_{4\sigma}\,  p_1\cdot p_3 \left[I_{\sigma\nu}\right]^R. 
\end{align}
This expression does vanish. The difference with~\refeq{nonzerodet} is that $\bar{\delta}_{\alpha\beta} k_\mu k_\nu \neq k^2$ if $k$ is a $d$-dimensional vector. Then,
\beq
\left[\bar{\delta}_{\alpha\beta} I_{\alpha\beta}\right]^R = \delta_{\alpha\beta} \left[ I_{\alpha\beta}\right]^R.
\eeq
Note that $\delta$ is the same as $\bar{\delta}$ outside $[.]^R$. We see that the rules in DReg are perfectly consistent in our example: $\text{Det}(\alpha\mu\nu;\beta\rho\sigma)$ does not vanish in $d$ dimensions while $\overline{\text{Det}}(\alpha\mu\nu;\beta\rho\sigma)$ can be safely set to zero in $n=2$. 

Things are very different in DRed. If we define $\widetilde{\text{Det}}$ just as in~\refeq{determinantbar} but with $\bar{\delta} \to \tilde{\delta}$, due to \refeq{projection} and the fact that the integration momentum $k$ is a $d$-dimensional vector (in the sense explained above), we find $\tilde{\delta}_{\alpha\beta} k_\mu k_\nu =  k^2$. Hence, we recover~\refeq{nonzerodet} and the inconsistency~\refeq{bad}. The root of the problem in this case is apparent: the relation~\refeq{projection} projects $n$-dimensional objects into $d$-dimensions, which invalidates the GnDI used for the former. 

Note that in DRed, the inconsistencies arise at the regularized level, due to the incompatibility of the dimensional reduction rule~\refeq{projection} with GnDI. In implicit methods, the GnDI are also dangerous before the identification and distinction of the different tensors. But they can be safely used afterwards: in CDR, after the (non-trivial) trace-traceless decompositions; in CIReg, after rewriting tensor integrals  and eliminating surface terms by generalizations of~\refeq{tensorCIR}; and in FDR, after the addition of $\mu^2$ in numerators, according to the global prescription. 

It will prove useful to mimic DReg and introduce in implicit methods a genuinely $n$-dimensional metric $\bar{\delta}$, with the properties\footnote{$\bar{\delta}$ and $\delta$ here play the same role as $\bar{\delta}$ and $\delta$, respectively, in DReg, except for the fact that in the latter method ${\delta}_{\mu\mu}=d$.}
\begin{align}
&\bar{\delta}_{\mu\nu} \bar{\delta}_{\nu\rho}  = \bar{\delta}_{\mu\nu} \delta_{\nu\rho} = \bar{\delta}_{\mu\rho}, ~~\text{(implicit)} \nn
& \bar{\delta}_{\mu\mu}=n.   \label{tdelta}
\end{align}
The distinguishing property of the metric $\bar{\delta}$ with respect to $\delta$ is that, by definition, 
\beq
\left[ \bar{\delta}_{\mu\nu}  T_{\dots\mu \dots \nu\dots}\right]^R = \delta_{\mu\nu} \left[T_{\dots\mu\dots \nu\dots}\right]^R, \label{bardeltaR}
\eeq
for any tensor $T$. In general, \refeq{bardeltaR} is different from $[T_{\dots \mu \dots \mu\dots}]^R$. In other words, for renormalization purposes $\bar{\delta}_{\mu\nu} k_\mu k_\nu= \bar{k}^2$ is not to be treated as $k^2$ but as if the indices were not contracted. For instance, in FDR, no $\mu^2$ is added to $\bar{k}^2$. (But once the $\mu^2$ shifts have been performed, one can write $\bar{k}^2=k^2$.) Because $\bar{\delta}$ commutes with renormalization, $\overline{\text{Det}}(\mu_1\mu_2\mu_3;\nu_1\nu_2\nu_3)$ vanish for $n=2$, just as in DReg. But importantly, in expressions such as~\refeq{fmunu}, it is still the ordinary metric $\delta$ of the formal $n$-dimensional space that appears. Otherwise, shift invariance or numerator-denominator consistency would be spoiled, as we have seen. If $E$ is either the $\epsilon$ tensor or an external tensor, then we can substitute at any moment one metric by the other one,
\beq
E_{\dots \mu \dots} \delta_{\mu \nu} = E_{\dots \mu \dots} \bar{\delta}_{\mu\nu},  \label{externaltilde}
\eeq
since the metrics appearing here will never contract two internal momenta, as long as GnDI are not employed. We can also use $\bar{\delta}$ in DRed, with the properties in~\refeq{tdelta} and~\refeq{projection}, supplemented with
\beq
\bar{\delta}_{\mu\nu} \tilde{\delta}_{\nu\rho} = \bar{\delta}_{\mu\rho} ~~~ \text{(DRed)}.
\eeq
In section~\ref{s:formal} we will comment on the structure of the spaces with these different metrics. 

The example we have examined may look quite contrived, but identities of this kind are often used to simplify expressions in the presence of completely antisymmetric tensors $\eps_{\mu_1\ldots \mu_n}$. This object can be defined formally by its rank and its antisymmetric character. Note that the definition is dimension-specific: even if we do not assign values to the indices, $\epsilon$ is only defined with $n$ indices. The  relations
\beq
\sum_{\pi\in S_{n+1}} \text{sign}(\pi) \eps_{\mu_{\pi(1)} \dots \mu_{\pi(n)}} \delta_{\mu_{\pi(n)} \nu} = 0 \label{Schouten}
\eeq
and
\beq
\epsilon_{\mu_1\dots\mu_n} \epsilon_{\nu_1\dots \nu_n} = \text{Det}(\mu_1\dots\mu_n ; \nu_1\dots \nu_n) \label{twoeps}
\eeq
are GnDI. They can lead to inconsistencies when used inside $[.]^R$. For instance, \refeq{Schouten} in $n=2$ implies
\begin{align}
0 & \overset{?}{=} \left[( \eps_{\mu\nu}\delta_{\rho \alpha} - \eps_{\mu\rho} \delta_{\nu \alpha} + \eps_{\nu\rho} \delta_{\mu\alpha}) I_{\mu\alpha} \right]^R   \nn
& = \pi \epsilon_{\rho\nu},  \label{nonzeroSchouten}
\end{align}
where we have used \refeq{Imunu}. One might be tempted to avoid some ambiguities by defining the left-hand side of~\refeq{twoeps} by its right-hand side, and in this way eliminate products of two $\eps$ tensors until one at most remains in a given factor. This definition is ill-defined. For instance, in a product $\eps_{\mu_1\nu_1} \eps_{\mu_2\nu_2} \eps_{\mu_3\nu_3} \eps_{\mu_4\nu_4}$,  it is possible to apply ~\refeq{twoeps} to three different pairs of pairs of $\eps$ tensors. The result with each choice is formally different and can give rise to different index contractions. Hence, when multiplied by a divergent integral, the result after renormalization may depend on how the four $\eps$ tensors have been paired. This is analogous to the DRed inconsistency pointed out in~\cite{Siegel2}.  On the other hand, no inconsistencies arise in any of the methods from GnDI such as~\refeq{Schouten} or~\refeq{twoeps} when the metric $\bar{\delta}$ is used instead of $\delta$.

\section{Dirac algebra}
\label{s:Dirac}
The Dirac matrices $\gamma_\mu$ transform as vectors under Lorentz transformations. In dimensional methods, they cannot have explicit $n$-dimensional form, since the Lorentz indices do not take explicit integer values. They are defined as a formal representation of the Clifford algebra:
\begin{align}
&\{ \gamma_\mu,{\gamma}_\nu \} = 2 {\delta}_{\mu\nu} \mathds{1}, ~~~~~\text{(DReg)} \label{dDirac}, \\
&\{ {\gamma}_\mu, {\gamma}_\nu \} = 2 \tilde{\delta}_{\mu\nu} \mathds{1} ,   ~~~~~\text{(DRed)} \label{nDirac} .
\end{align}
Trace identities follow in each case from these definitions, the ciclicity of the trace (which we assume throughout the paper) and the value of the trace of the identity, which in both methods can be taken to be $\mathrm{tr} \mathds{1} = n$. Because of the projection rule~\refeq{projection}, even if the Dirac algebra looks $n$-dimensional in DRed, this can be effectively changed by contractions with the integration momenta. In fact, the relation $\kslash \! \kslash \! = k^2$ is
 necessary to preserve numerator-denominator consistency. 
Implicit methods also treat Lorentz tensors in a formal way, as we have seen, so the Dirac matrices are naturally defined by
\beq
\{{\gamma}_\mu,{\gamma}_\nu \} = 2 {\delta}_{\mu\nu} \mathds{1},  ~~~~~\text{(implicit)}  \label{Dirac} ,
\eeq
where of course $\delta$ here is (formally) $n$-dimensional. 

The formal treatment of the Dirac algebra in all of these methods suffers from a fundamental problem when $n$ is odd. This can be checked most easily in DReg~\cite{Anselmi}. First,~\refeq{dDirac} and the cyclicity of the trace imply
\beq
d \text{tr}(\gamma _\mu) = \text{tr}(\gamma _\mu \gamma _\alpha \gamma _\alpha) = \text{tr}(\gamma _\alpha \gamma _\mu \gamma _\alpha) 
= 2 \text{tr}(\gamma _\mu)-\text{tr}(\gamma _\mu \gamma _\alpha \gamma _\alpha) = (2-d) \text{tr}(\gamma _\mu),
\eeq
Hence, unless $d=1$, $\text{tr}(\gamma _{\mu})=0$. Similar manipulations for a product of an odd number $m$ of Dirac matrices lead to
\beq
(d-m) \tr(\gamma _{\mu_1} \dots \gamma _{\mu_m}) = 0.
\eeq
Therefore, $\tr(\gamma _{\mu_1} \dots \gamma _{\mu_m}) = 0$ unless $d=m$. Analytical continuation in $d$ then requires all these products to vanish identically for all $d$. But this is incompatible with the fact that the product of $n$ Dirac matrices in $n$ fixed odd dimensions is proportional to the $\eps$ tensor, a property that should be recovered after renormalization. To solve this problem, one must break the $d$-dimensional Lorentz covariance of the Dirac algebra changing $\delta$ by $\bar{\delta}$ in~\refeq{dDirac}, as proposed in~\cite{Anselmi}. This is consistent with~\refeq{projectionDReg} but compromises numerator-denominator consistency. On the other hand, even if the definition~\refeq{nDirac} is employed in DRed, the problem reappears when the indices in the initial trace are contracted with integration momenta, due to the projection rule~\refeq{projection}. 

Presented in this way, the inconsistency in odd dimensions looks like a specific problem of the analytical continuation in $d$. However, it turns out that implicit methods also treat the Dirac algebra inconsistently when the dimension $n$ is odd. Let us show it for the case $n=3$, for definiteness. In three dimensions,\footnote{Lorentz covariance guarantees that this trace is proportional to the $\eps$ tensor. The numerical factor can be determined by agreement with the usual algebraic result, for instance using the Pauli matrices as a representation of the 3-dimensional Dirac algebra.} 
\beq
\text{tr}(\gamma_{\mu} \gamma_{\nu} \gamma_{\rho}) = 2\eps_{\mu\nu\rho}.
\eeq
Then, from~\refeq{nDirac} and the cyclicity of the trace,
\begin{align}
\tr(\ga_\mu\ga_\nu\ga_\rho\ga_\sigma\ga_\tau) & = - \tr(\ga_\nu \ga_\mu \ga_\rho\ga_\sigma\ga_\tau) + 2\delta_{\mu\nu} \tr(\ga_\rho\ga_\sigma\ga_\tau) \nn
& =  \tr(\ga_\mu\ga_\nu\ga_\rho\ga_\sigma\ga_\tau) + 2 \left(\delta_{\mu\nu} \tr(\ga_\rho\ga_\sigma\ga_\tau) - \delta_{\mu\rho} \tr(\ga_\nu\ga_\sigma\ga_\tau) +  \delta_{\mu\sigma}\tr(\ga_\nu\ga_\rho\ga_\tau) \right. \nn
& \phantom{=} \left. \mbox{} -  \delta_{\mu\tau}\tr(\ga_\nu\ga_\rho\ga_\sigma)\right) \nn
& = \tr(\ga_\mu\ga_\nu\ga_\rho\ga_\sigma\ga_\tau) + 4 \left(\delta_{\mu\nu} \eps_{\rho\sigma\tau} - \delta_{\mu\rho} \eps_{\nu\sigma\tau} +  \delta_{\mu\sigma}\eps_{\nu\rho\tau} - \delta_{\mu\tau}\eps_{\nu\rho\sigma} \right),
\end{align}
which in view of~\refeq{Schouten} looks fine at first sight. However, as we have seen in the previous section the combination of $\eps$ tensors and deltas in the last line needs not vanish inside~$[.]^R$ when two of the indices are contracted with the integration momenta of a divergent integral. Therefore, the result of the calculations can be ambiguous. 

From now on, we will assume that the dimension $n$ is even, unless otherwise indicated.  One of the most important limitations of not being able to employ GnDI is the absence of a finite complete set in Dirac space. In ordinary $n$-dimensional space, the antisymmetric products 
\begin{align}
[\mu_1\dots \mu_m] & = \frac{1}{m!} \sum_{\pi\in S_{m}} \text{sign}(\pi)  \gamma_{\mu_{\pi(1)}} \cdots \gamma_{\mu_{\pi(m)}}, ~~~ m=1,\ldots ,n,
\end{align}
together with the identity $\mathds{1}$, form a linearly independent complete set of the space of $2^{n/2}\times2^{n/2}$ complex matrices.\footnote{We are discussing the case of even $n$. For odd $n$, the set $\{[\mu_1\dots \mu_m], m=1,\dots (n-1)/2 \}$ is a complete set of $2^{(n-1)/2}\times2^{(n-1)/2}$ matrices.} In the formal $n$-dimensional space, the Dirac matrices cannot be understood as matrices of any specific dimension, so completeness must be defined also in a formal way.  As shown in~\cite{gammatrica}, many useful relations can be proven using only formal manipulations. The matrices $[\mu_1 \dots \mu_m]$ are orthogonal with respect to the trace bilinear form. Then, a string of Dirac gamma matrices
\beq
S_{\alpha_1 \dots \alpha_m} = \gamma_{\alpha_1} \dots \gamma_{\alpha_m} \label{string}
\eeq
can always be written as
\beq
S_{\alpha_1 \dots \alpha_m} = a^{\alpha_1 \dots \alpha_m}_0 \mathds{1} + a^{\alpha_1 \dots \alpha_m}_\mu [\mu] + \cdots + a^{\alpha_1 \dots \alpha_m}_{\mu_1  \dots \mu_m} [\mu_1  \dots \mu_m] \label{complete},
\eeq
with ($n$-independent) coefficients given by
\beq
a^{\alpha_1 \dots \alpha_m}_{\mu_1  \dots \mu_k} = \frac{1}{n m!} \tr \left( S_{\alpha_1 \dots \alpha_m} [\mu_k  \dots \mu_1] \right) . \label{coefficients}
\eeq
Therefore, $\mathcal{B}=\{\mathds{1},[\mu_1],[\mu_1\mu_2], \dots \}$ is a countable Hamel basis of the formal Dirac space, defined as the set of arbitrary linear combinations of strings of the form~\refeq{string} (including the case with $m=0$, $S=\mathds{1}$). The main difference with a genuine $n$-dimensional space is that the objects $[\mu_1 \dots \mu_m]$ do not vanish for $m>n$, so the space is infinite-dimensional. For instance, in formal $n$-dimensional space we have
\beq
S_{\mu\nu\rho} = \delta_{\mu\nu} \gamma_\rho - \delta_{\mu\rho} \gamma_\nu+\delta_{\nu\rho} \gamma_\mu + [\mu\nu\rho], 
\eeq
which is valid for any even $n$, including $n=2$. Using the mentioned GnDI, in $n=2$ we could instead simplify this expression to
\beq
S_{\mu\nu\rho} = \delta_{\mu\nu} \gamma_\rho - \delta_{\mu\rho} \gamma_\nu+\delta_{\nu\rho} \gamma_\mu.
\eeq
But as stressed already many times, such simplifications are dangerous before renormalization.

The standard Fierz identities in $n$ dimensions can be derived using the completeness of $\{\mathds{1},[\mu_1],\dots, [\mu_1 \dots \mu_n]\}$. Similarly, in the formal $n$-dimensional space one can derive Fierz identities from the completeness of $\mathcal{B}$. However, the Fierz reorderings in this case involve in general an infinite number of terms, just as in DReg~\cite{Fierzdimensional}, which makes them less useful. Moreover, the invariance under supersymmetry transformations of the action of supersymmetric theories relies on genuine $n$ dimensional Fierz identities (and also on an anticommuting $\gamma_5$).  In fact, as shown in~\cite{Avdeev,Avdeev2,Stoeckinger}, the supersymmetry Ward identities are violated when relevant GnDI are not fulfilled.

In even dimensions, Weyl spinors can be defined from Dirac spinors by chiral projectors constructed with $\gamma_5$.\footnote{We call this object $\gamma_5$ for any integer dimension $n$.  Because in this paper we never write Lorentz indices with explicit integer values, no confusion with $\gamma_\mu$ should arise.} Several definitions of $\gamma_5$ are in principle possible in the methods we are considering. First, it can be defined formally by the basic property
\beq
\{\gamma_5^{\text{AC}},\gamma_\mu\} = 0 , \label{anticommuting}
\eeq
where the label AC has been introduced to distinguish this definition from the one we favor below. This simple definition is consistent, as has been proven in~\cite{Stoeckinger} by explicit construction. Unfortunately, in all the methods we consider, it is incompatible with the correct $n$-dimensional value of odd-parity traces.
This fact is well known in DReg~\cite{BM,Bonneau}. In $n=2$, for example, after renormalization we would like to recover the standard value
\beq
\text{tr}(\gamma_5 \gamma_\mu\gamma_\nu) = -2 \epsilon_{\mu\nu} .  \label{2Dtrace}
\eeq
On the other hand, using \refeq{anticommuting} and the DReg rules in $\text{tr}(\gamma^{\text{AC}}_5 \gamma_\mu\gamma_\nu \gamma_\rho \gamma_\rho)$, it follows that 
\beq
d(d-2) \text{tr}(\gamma^{\text{AC}}_5 \gamma_\mu \gamma_\nu) = 0,
\eeq
which shows that $\text{tr}(\gamma^{\text{AC}}_5 \gamma_\mu \gamma_\nu)$ vanishes identically and \refeq{2Dtrace} cannot be recovered in the limit $d\to 2$. DRed faces the same situation when the free indices in the initial trace are contracted with integration momenta, due to the projection rule~\refeq{projection}~\cite{Bonneau}. Once again, this issue appears as well in implicit methods. Indeed,~\refeq{anticommuting} and~\refeq{2Dtrace} imply
\beq
\tr(\gamma^{\text{AC}}_5 \gamma_\mu \gamma_\nu \gamma_\rho \gamma_\sigma) =  \tr(\gamma^{\text{AC}}_5 \gamma_\mu \gamma_\nu \gamma_\rho \gamma_\sigma) - 2 \eps_{\nu\rho} \delta_{\mu\sigma} + 2 \eps_{\mu\rho} \delta_{\nu\sigma} - 2 \eps_{\mu\nu} \delta_{\rho\sigma} .
\eeq
Again, in spite of GnDI~\refeq{Schouten} and as shown in~\refeq{nonzeroSchouten}, $- \eps_{\nu\rho}\delta_{\mu\rho} + \eps_{\mu\rho} \delta_{\nu\sigma} - \eps_{\mu\nu} \delta_{\rho\sigma}$ can be nonzero inside $[.]^R$, which then leads to a contradiction.\footnote{This argument in even dimensions is almost identical to the one above in odd dimensions. Taking into account that the usual candidate for $\gamma_5$ is proportional to the identity in odd dimensions, we see that the origin of the inconsistencies is essentially the same in odd and even dimensions.}  In the same way it can be shown that $\text{tr}(\gamma^{\text{AC}}_5 \gamma_{\mu_1} \dots \gamma_{\mu_n})=0$ for any even $n$.  This is certainly not what one would want in an $n$-dimensional method and it shows that the definition~\refeq{anticommuting} does not provide a correct regularization of arbitrary diagrams in a chiral theory.  Note that other traces with one $\gamma^{\text{AC}}_5$ matrix also vanish, since they must be antisymmetric and there is no Lorentz-covariant completely antisymmetric tensor of rank $m\neq n$. This can be extended to traces with an odd number of $\gamma^{\text{AC}}_5$ matrices if $(\gamma^{\text{AC}}_5)^2=-\mathds{1}$, a property which is required to form chiral projectors.

This problem of $\gamma^{\text{AC}}_5$ reappears in a more subtle form in open fermion lines. To see this, assume for a moment that $\gamma^{\text{AC}}_5$ belongs to the formal Dirac space, as defined above. Then, using the completeness of $\mathcal{B}$, we would find
\begin{align}
\gamma^{\text{AC}}_5 & = \frac{1}{n} \text{tr} (\gamma^{\text{AC}}_5) \mathds{1}+ \frac{1}{2 n} \text{tr} ( \gamma^{\text{AC}}_5 [\nu \mu]) [\mu \nu] + \frac{1}{24n} \text{tr} (\gamma^{\text{AC}}_5 [\sigma \rho \nu \mu]) [\mu \nu \rho \sigma] + \dots \nn
&= 0.
\end{align}
Therefore, if $\gamma^{\text{AC}}_5$ is to be nontrivial, it cannot belong to the formal Dirac space.\footnote{This is apparent in the explicit construction of~\cite{Stoeckinger}.} But then, the eventual projection into the standard Dirac space of genuine $n$-dimensional space, which is a subset of the former, will annihilate it. So, to recover standard Dirac strings with $\gamma_5$ matrices, one needs to replace by hand $\gamma^{\text{AC}}_5$ by $\gamma_5$ after renormalization. It does not seem obvious to us that this {\em ad hoc} replacement in multiloop amplitudes will respect unitarity.    

An alternative definition of $\gamma_5$ is to generalize its explicit definition in genuine $n$ dimensions in terms of the Dirac matrices:
\beq
\gamma_5 = \frac{1}{n!} \eps_{\mu_1 \ldots \mu_n} \gamma_{\mu_1} \cdots \gamma_{\mu_n} . \label{gamma5def}
\eeq
This is akin to the original tHV definition in DReg~\cite{tHV} and is the definition we will use in the following, unless otherwise indicated.
Note that, even if we are not restricting the indices to have $n$ different values, this object is $n$-dimensional in the sense that it contains $n$ Dirac matrices. Furthermore, in view of~\refeq{externaltilde}, we can write~\refeq{gamma5def} in the alternative form
\beq
\gamma_5 = \frac{1}{n!} \eps_{\mu_1 \ldots \mu_n} \bar{\gamma}_{\mu_1} \cdots \bar{\gamma}_{\mu_n} ,
\eeq
where $\bar{\gamma}_\mu = \bar{\delta}_{\mu\nu} \ga_\nu$. Like any other explicit definition, \refeq{gamma5def} does not introduce any consistency issues by itself. The non-trivial question is which familiar properties of the $\gamma_5$ can be proven without using dangerous GnDI. 
The most important of these properties is the anticommutation with the Dirac matrices, but from the discussion above it is clear that this property cannot hold for the definition~\refeq{gamma5def} in any of the methods we are discussing.\footnote{The fact that $\gamma_5$ does not anticommute with the Dirac matrices has already been observed in FDR~\cite{FDR} and CIReg~\cite{CIRanomaliesI,CIRanomaliesII}.} 
Indeed, for $n=2$, for instance, \refeq{gamma5def} and~\refeq{nDirac} give
\beq
\tr(\ga_5 \ga_\mu \ga_\nu \ga_\rho \ga_\sigma) + \tr(\ga_\mu \ga_5 \ga_\nu \ga_\rho \ga_\sigma) = -4\left(\delta_{\mu\nu}\eps_{\rho\sigma}-\delta_{\mu\rho} \eps_{\nu\sigma} + \delta_{\mu\sigma} \eps_{\nu\rho}  \right)
\eeq
This expression vanishes when it accompanies finite integrals. However, using~\refeq{nonzeroSchouten} we get
\beq
\left[ \left(\tr(\ga_5 \ga_\mu \ga_\nu \ga_\rho \ga_\sigma) + \tr(\ga_\mu \ga_5 \ga_\nu \ga_\rho \ga_\sigma)\right) I_{\mu\sigma} \right]^R =4 \pi \eps_{\rho\nu}
\eeq 
In the same vein, let us point out that some of the explicit trace expressions of odd-parity products of Dirac matrices in the literature have been simplified with the help of the GnDI~\refeq{Schouten}. To avoid inconsistencies,  only the complete expressions derived from \refeq{gamma5def} and~\refeq{nDirac} or~\refeq{Dirac} should be used before renormalization. The nonvanishing anticommutator $\{\ga_5,\ga_\mu\}$ can be written in a simple form using $\bar{\delta}$. First, observe that in $n=2$,
\begin{align}
0 & = (\gamma_\mu\gamma_\nu\gamma_\rho) (\eps_{\mu\nu} \bar{\delta}_{\rho\alpha} - \eps_{\mu\rho}\bar{\delta}_{\nu\alpha}+\eps_{\nu\rho}\bar{\delta}_{\mu\alpha}) \nn
& = - \eps_{\mu\rho}(\ga_\mu \ga_\rho \bar{\ga}_\alpha+\bar{\ga}_\alpha \ga_\rho \ga_\mu) \nn
& = \{\gamma_5,\bar{\gamma}_\alpha\}. \label{antig5}
\end{align}  
From this, similarly to DReg, we find
\beq
\{\gamma_5,\gamma_\alpha\} =  2 \ga_5 \hat{\ga}_\alpha, \label{anticonmutador}
\eeq
where we have introduced the evanescent metric $\hat{\delta}=\delta - \bar{\delta}$, which has trace $\hat{\delta}_{\mu\mu}=0$, to write the evanescent matrix $\hat{\ga}_\mu = \hat{\delta}_{\mu\nu} \ga_\nu=\ga_\mu - \bar{\ga}_\mu$, and used the fact that this matrix commutes with $\gamma_5$. Indeed, in $n=2$,
\begin{align}
[\ga_5,\hat{\ga}_\alpha] & = [\ga_5,{\ga}_\beta] (\delta_{\beta \alpha}-\bar{\delta}_{\beta\alpha}) \nn 
& = \frac{1}{2} \eps_{\mu\nu} [\ga_\mu \ga_\nu,\ga_\beta] (\delta_{\beta \alpha}-\bar{\delta}_{\beta\alpha}) \nn 
& = -2 \epsilon_{\beta \mu}\gamma_\mu (\delta_{\beta \alpha}-\bar{\delta}_{\beta\alpha}) \nn
& = 0,
\end{align}
due to \refeq{externaltilde}. The proof of~\refeq{anticonmutador} can be generalized to arbitrary even $n$. Let us also note in passing the useful relations
\begin{align}
& \{\bar{\ga}_\mu,\bar{\ga}_\nu\} =  \{\bar{\ga}_\mu,\ga_\nu\} =2 \bar{\delta}_{\mu\nu}, \nn
& \{\hat{\ga}_\mu,\hat{\ga}_\nu\} =  \{\hat{\ga}_\mu,{\ga}_\nu\} = 2 \hat{\delta}_{\mu\nu}, \nn
& \{\bar{\ga}_\mu,\hat{\ga}_\nu\} = 0,
\end{align}
which follow from the definitions of the involved objects.
Similarly, $\bar{\delta}$ can be used to show that $\gamma_5^2=-\mathds{1}$ in any even $n$. In $n=2$, for example,
\begin{align}
\gamma_5 \gamma_5 & = \frac{1}{4} \eps_{\mu\nu} \eps_{\rho\sigma} \ga_\mu \ga_\nu \ga_\rho \ga_\sigma \nn
& = \frac{1}{4} (\bar{\delta}_{\mu\rho}\bar{\delta}_{\nu\sigma}-\bar{\delta}_{\mu\sigma}\bar{\delta}_{\nu\rho}) \ga_\mu \ga_\nu \ga_\rho \ga_\sigma \nn
& = -\mathds{1}.
\end{align}
In the second line we have used the GnDI~\refeq{twoeps}, involving only the $\epsilon$ tensors. In the last one,~\refeq{nDirac} and~\refeq{tdelta}.

\section{A consistent procedure in implicit fixed-dimension methods}
\label{s:formal}
In even dimension $n$, the inconsistencies of DRed can be avoided simply by forbidding the use of GnDI before renormalization, as proposed in~\cite{Avdeev,Avdeev2}. That is, the $n$-dimensional space to be used in a consistent version of DRed is not the genuine $n$-dimensional Euclidean space (GnS), but a quasi-$n$-dimensional space (QnS).  Similarly to the case of quasi-$d$-dimensional space (QdS) in DReg~\cite{Collins}, QnS can be defined explicitly as an infinite-dimensional vector space endowed with a metric $\tilde{\delta}$, which satisfies $\tilde{\delta}_{\mu\mu}=n$~\cite{Stoeckinger}. The relation with QdS is given by the direct-sum structure $\text{QnS=QdS} \oplus \text{Q}\varepsilon\text{S}$. Dirac matrices in the three spaces have been explicitly constructed in~\cite{Stoeckinger}, following~\cite{Collins}.

We propose here to define implicit methods in the same~QnS. In this case, there is no need to embed QdS in it, so the setup is simpler. Moreover, the metric can be called $\delta$ without confusion, in agreement with our notation thus far.  Forbidding GnDI is actually not sufficient in fixed dimension, since the discrimination of Lorentz tensors is not automatic. As anticipated above, we need to specify some normal form of the expressions to uniquely identify the different tensor structures.\footnote{As a matter of fact, some standard form is also required in the dimensional methods to display explicitly all the $d$ dependence and thus be able to apply MS or $\overline{\text{MS}}$ without ambiguities.} 
Following~\cite{BM}, we propose to simplify arbitrary Feynman diagrams with the following algorithm, which leads to a unique normal form:
\begin{description}
\item{(i)} All $\gamma_5$ are substituted by their tHV definition~\refeq{gamma5def}.
\item{(ii)} All Dirac matrices are removed from denominators.
\item{(iii)} Dirac traces are computed using $\tr AB = \tr BA$, \refeq{nDirac} and $\tr \mathds{1}=n$.
\item{(iv)} Products of Dirac gammas are decomposed into sums of antisymmetric combinations as in~\refeq{complete} and~\refeq{coefficients}.
\item{(v)}  All possible contractions are performed, using $\delta_{\mu\nu} V_{\dots \nu \dots} \to V_{\dots \mu \dots}$ for arbitrary tensors $V$.
\item{(vi)} $\delta_{\mu \mu}$ is replaced by $n$.
\end{description}
As we work in QnS from the start, GnDI cannot be applied. Indeed, if GnDI were allowed, the resulting expression would not have unique form, which could eventually translate into different renormalized results. There are however exceptions to this prohibition, which are discussed below. After performing the algebraic manipulations in steps (i--vi), the diagram will be a sum of terms that contain $\eps$ tensors, metrics with free indices, antisymmetric arrays of gamma functions, external momenta, possible background tensors and a tensor (multi-dimensional) integral $T$. In this way, the different integrals $T$ that appear in a given diagram are determined. They are then to be renormalized as prescribed in the different methods. After this, there is no harm in using GnDI. In particular, they can and should be used after subtraction to simplify the final results. Note in particular that, because the final antisymmetric combinations of Dirac matrices $[\alpha_1 \ldots \alpha_m]$ are not touched by renormalization, only the combinations with $m\leq n$ need to be included in the decomposition of step (iv).

Sometimes selected GnDI can be used to simplify expressions from the very beginning, as long as one is sure that they will not change the contractions of indices in the loop integrals $T$. One simple example in $n=2$ is using $\eps_{\mu\nu}\eps_{\mu\nu}=2$.
More generally, we can simplify the calculations significantly using the metric $\bar{\delta}$, defined above. The rules it obeys,~\refeq{tdelta}, can be understood as the consequence of the structure $\text{QnS}=\text{GnS}\oplus \text{X}$, with $\delta$, $\bar{\delta}$ and $\hat{\delta}=\delta-\bar{\delta}$ the metrics in QnS, GnS and the extra space X, respectively. Remember that the defining property of $\bar{\delta}$ in implicit methods is that it commutes with renormalization. 
In expressions related to loop integrals, such as~\refeq{fmunu}, or in the traces of Dirac matrices, it is still the ordinary metric $\delta$ of QnS that appears, to comply with shift invariance and numerator-denominator consistency.  The idea here is to allow for GnDI that involve {\em only\/} $\bar{\delta}$, the $\epsilon$ tensor and external momenta or fields. Then, $\bar{\delta}$ can appear as the result of these GnDI.  
Using such GnDI spoils the uniqueness of the normal form. However, the resulting expressions have the same renormalized value, thanks to~\refeq{bardeltaR}. As a straightforward illustration in $n=2$,
\begin{align}
\left[ \eps_{\mu\nu} \eps_{\nu\rho} I_{\mu\rho}\right]^R & = \eps_{\mu\nu} \eps_{\nu\rho} \left[I_{\mu\rho}\right]^R \nn
& = 2 \delta_{\mu\rho} \left[I_{\mu\rho}\right]^R \nn
& = \left[ 2 \bar{\delta}_{\mu\rho} I_{\mu\rho} \right]^R .
\end{align}
In the next section we give simple examples that illustrate how the calculations can be simplified with the help of $\bar{\delta}$ and related objects.

The same simplifications are valid also in the consistent version of DRed~\cite{Stoeckinger} with a tHV $\gamma_5$. The only difference is that in this method four different spaces are used, related by $\text{QnS=QdS}\oplus \text{Q}\varepsilon \text{S}$ and $\text{QdS=GnS} \oplus \text{Q}(-\varepsilon)\text{S}$. Then, we can identify the extra space in fixed dimension with $X=\text{Q}(-\varepsilon)\text{S}\oplus \text{Q}\varepsilon \text{S}$. The relations between the metrics in~\refeq{projection} and \refeq{tdelta} are those implied by this hierarchical structure, with $\tilde{\delta}$, $\delta$ and $\bar{\delta}$ the metrics in QnS, QdS and GnS, respectively. 

We have already pointed out that GnDI can be safely used after tensor identification. Indeed, after that step, $\delta$ behaves as $\bar{\delta}$. This is specially relevant to FDR, as in this method some useful shortcuts exist to identify tensors from the very start. As a salient example, in one-loop diagrams with fermion lines that do not have indices contracted with the ones in other fermion lines, it is easy to see that the correct $\mu^2$ shifts can be obtained by shifting (in Euclidean space) the integration momenta as $1/\kslash \to 1/(\kslash \pm i \mu)$, with opposite signs for $\kslash$ separated by an even number of  $\gamma$ matrices and equal signs for those separated by an odd number of $\gamma$ matrices. For this, it is important that terms with odd powers of $\mu$ do not contribute after the limit $\mu\to 0$. We can easily generalize this rule to spinor chains that contain $\gamma_5$ matrices: because, according to its definition~\refeq{gamma5def}, $\gamma_5$ contains an even number of $\gamma$ matrices in even dimension, the $\gamma_5$ matrices should just be ignored in the determination of the signs. This approach allows, for instance, to use an anticommuting $\gamma_5$ before evaluating Dirac traces. The results are unique and agree with the ones obtained from the normal form or with the $\bar{\delta}$ formalism. When one Lorentz index is contracted between different fermion lines, a similar, more complicated rule can be found which gives the right $\mu^2$~\cite{Alicethesis}. Modifications may also be necessary in diagrams that contain both Dirac traces and derivative interactions. To the best of our knowledge, no general prescription exists to treat any diagram in this way.  A very similar idea is used in~FDF. In this dimensional method, the necessary $\mu^2$ are obtained from the extra-dimensional components of integration momenta and a set of selection rules for the extra-dimensional space (see also~\cite{PittauPrimary}). Then, GnDI are valid and $\gamma_5$ anticommutes with the Dirac matrices. Comparing with the situation in FDR, it seems that in order to comply with the quantum action principle the method will require some refinements for multiloop calculations. 

The consistent procedure for implicit methods in $\text{QnS}$ can in principle be applied to multi-loop calculations. A careful rigorous discussion goes well beyond the scope of this paper, but let us sketch how the renormalization  of a Feynman diagram could proceed. First, the diagram is expressed in its normal form, following the steps above. Allowed GnDI involving $\bar{\delta}$ can be optionally used. Then, each tensor integral $T$ is treated with Bogoliubov's recursive $R$-operation~\cite{Bogoliubov:1957gp,Hepp:1966eg} (or equivalently its solution, Zimmermann's forest formula~\cite{forest}), in order to guarantee locality and unitarity of the renormalized theory.  To do this, a subtraction operator, which selects the singular part of a primitively divergent (sub)graph $\Gamma$ of $T$, can be defined without any explicit regularization as $K \Gamma = \tilde{R} \Gamma - \Gamma$~\cite{forestCDR}. Here, $\tilde{R} \Gamma$ is $\Gamma$ with its (overall) divergence subtracted. Then, $K$ is applied according to Bogoliubov's formula. 

This systematic method has been used in differential renormalization~\cite{forestCDR} and in CIReg~\cite{forestCIR}, but only in non-derivative scalar theories, which have a simple tensor structure. In more complicated theories, it is essential to treat tensor integrals consistently. To do this, in calculating $\tilde{R} \Gamma$ for a tensor $\Gamma\subset T$, the Lorentz indices in $\Gamma$ that are contracted with indices in $T\backslash \Gamma$ should be treated as uncontracted free indices. This is a necessary condition to preserve invariance under shifts of the integration momenta in $\Gamma$ that are proportional to the integration momenta in $T\backslash \Gamma$.  We will not try to prove here that it is also a sufficient condition for shift invariance of the final renormalized multi-loop integrals. This issue has been addressed in particular examples in CIReg~\cite{Dias:2008iz} and FDR~\cite{FDRloops}. We believe that the so-called {\em extra-extra integrals} that are introduced in FDR to impose sub-integration consistency are equivalent to the contribution of (sums of) forests with the tensor rule above. They are also related to the DRed contributions of $\varepsilon$ scalars associated to virtual vector bosons, which renormalize independently.

Finally, we should stress that, even if implicit methods as treated in this section are consistent and preserve shift invariance and numerator-denominator consistency, some particular Ward identities based on GnDI may be broken. This is the origin of chiral anomalies and of the breakdown of supersymmetry. Also vectorial Ward identities associated to gauge invariance can be broken in the presence of the tHV $\gamma_5$, giving rise to spurious anomalies that must be eliminated with additional finite counterterms. We will give an example of this in the next section. In this regard, these methods are not better or worse than DReg.

\section{Examples}
\label{s:examples}
We will present simple off-shell calculations for non-exceptional momenta in the Euclidean region, such that no infrared divergences can arise. 
\subsection{Vector and axial currents in two dimensions}
Let us consider a free massless Dirac fermion in Euclidean space of dimension $n=2$, with Lagrangian
\beq
\mathcal{L} = \bar{\psi}  \slashed{\d} \psi .  \label{freeLag}  
\eeq
This Lagrangian is invariant under global vector (V) and axial (A) transformations. The corresponding, classically conserved Noether currents are
\begin{align}
& j_\mu = \bar{\psi} \ga_\mu \psi ,\\
& j^5_\mu = \bar{\psi} \ga_\mu \ga_5 \psi,
\end{align}
respectively.\footnote{Because this current will always be an external operator in our calculations, nothing would change should we write instead $j^5_\mu = - \bar{\psi} \ga_5 \ga_\mu \psi$ or the average of these two definitions.} We want to calculate the correlation functions of two of these currents. The three distinct possibilities are $\Pi_{\mu\nu}(p) = \langle j_\mu(p) j_\nu(-p) \rangle$, $\Pi^5_{\mu\nu}(p) = \langle j_\mu(p) j^5_\nu(-p) \rangle$ and $\Pi^{55}_{\mu\nu}(p) = \langle j^5_\mu(p) j^5_\nu(-p) \rangle$. The classical Ward identities are
\begin{align}
& p_\mu \Pi_{\mu\nu}(p) = p_\nu \Pi_{\mu\nu}(p) = 0, \label{vector} \\
& p_\mu \Pi^5_{\mu\nu}(p) = 0, \label{vectoraxial} \\
& p_\nu \Pi^5_{\mu\nu}(p) = 0, \label{clasaxial} \\
& p_\mu \Pi^{55}_{\mu\nu}(p) = p_\nu \Pi^{55}_{\mu\nu}(p) = 0, \label{clasaxialaxial}
\end{align}
A useful GnDI in $n=2$ is $\gamma_\mu \gamma_5 = \epsilon_{\mu\alpha} \gamma_\alpha$.  This can be proven, for instance, using the complete set in GnS Dirac space. The correlation functions can be calculated exactly at one loop. Before doing it, we can anticipate the form of the correlators. In fact, the previous GnDI implies $j^5_\mu = \epsilon_{\mu\alpha} j_\alpha$, so the three correlators are algebraically related:
\begin{align}
& \Pi^5_{\mu\nu}(p) = \epsilon_{\nu_\alpha} \Pi_{\mu\alpha} , \label{simple5} \\
& \Pi^{55}_{\mu\nu}(p) = \delta_{\mu\nu} \Pi_{\alpha\alpha} - \Pi_{\nu\mu} \label{simple55} .
\end{align}
In the second of these equations we have also used the GnDI~\refeq{twoeps} for $n=2$. From this, 
we can easily conclude that the Ward identities~(\ref{vector}--\ref{clasaxialaxial}) cannot be satisfied simultaneously. Indeed, dimensional analysis and the fact that the longitudinal piece is finite imply
\beq
\Pi_{\mu\nu}(p) = X \left( \frac{p_\mu p_\nu}{p^2} - a \delta_{\mu\nu} \right), \label{guess}
\eeq
where both $X$ and $a$ are numbers. $X$ is fixed by the result of a finite integral, while $a$ is regularization dependent and can be modified with a local finite counterterm.  In order to fulfill~\refeq{vector}, we need $a=1$. Then,  we see that~\refeq{vectoraxial} is also satisfied but~\refeq{clasaxial} and~\refeq{clasaxialaxial} are not. Instead, we have the anomalous identities
\begin{align}
& p_\nu \Pi^5_{\mu\nu}(p) =  X \epsilon_{\mu\nu} p_\nu , \label{qaxial} \\
& p_\mu \Pi^{55}_{\mu\nu}(p) = - X p_\nu . 
\end{align}
It should be noted that all the GnDI we have employed involve external tensors only. Therefore, we expect that these results hold in consistent regularization and renormalization schemes that respect~\refeq{vector}, including the method proposed in the previous section.

Let us now check this by explicit computation. We will use FDR for definiteness and because it allows us to compare with the rule that allows to identify the tensor integrals a priori, before computing the trace. We have checked that all the results are identical in CDR and CIReg and also in consistent DRed and FDF. Because no $\delta_{\alpha\alpha}$ arises from the Dirac matrices, the results in DReg are identical as well in these examples. The only contributing diagram to the VV correlator gives
\beq
\Pi_{\mu\nu}(p)=- \left[\int \frac{d^2 k}{4\pi^2} \, \text{tr} \left(\gamma_{\mu}\frac{1}{\slashed{k}-\slashed{p}}\gamma_{\nu}\frac{1}{\slashed{k}} \right) \right]^R.
\eeq
Performing the trace, we find
\beq
\Pi_{\mu\nu}(p)=- \left[4 B_{\mu\nu}(p) - 2 \delta_{\mu\nu} B_{\alpha \alpha}(p) \right]^R , \label{pimununormal}
\eeq
where
\beq
B_{\alpha\beta}(p) = \int \frac{d^2 k}{4\pi^2} \, \frac{(k-p)_\alpha k_\beta}{(k-p)^2 k^2}.
\eeq
Note that~\refeq{pimununormal} is written in normal form.
In FDR, we have
\begin{align}
\left[B_{\alpha\beta}(p)\right]^R & = \left[ \int \frac{d^2 k}{4\pi^2} \, \frac{(k-p)_\alpha k_\beta}{[(k-p)^2+\mu^2] [k^2+\mu^2]} \right]^S \nn
& = \frac{1}{4\pi} \left\{\delta_{\alpha\beta} \left(1-\frac{1}{2} \log \frac{p^2}{\mu^2} \right) - \frac{p_\alpha p_\beta}{p^2} \right\} , \label{Bmunu}
\end{align}
whereas
\begin{align}
\left[B_{\alpha\alpha}(p)\right]^R & = \left[ \int \frac{d^2 k}{4\pi^2} \, \frac{(k-p)_\alpha k_\alpha + \mu^2}{[(k-p)^2+\mu^2] [k^2+\mu^2]} \right]^S \nn
& = \delta_{\alpha\beta} \left[ B_{\alpha\beta}(p)\right]^R - \frac{1}{4\pi} \nn
& = - \frac{1}{4\pi} \log \frac{p^2}{\mu^2} . \label{Balal}
\end{align}
The extra local term in the second equality comes, just as in~\refeq{tensorFDR}, from the oversubtracted integral proportional to $\mu^2$, which is added to the numerator in the first line, according to the global prescription. Combining everything, we find
\beq
\Pi_{\mu\nu}(p)=\frac{1}{\pi} \left(\frac{p_\mu p_\nu}{p^2} - \delta_{\mu\nu}\right), 
\eeq
which agrees with~\refeq{guess} with $X=1/\pi$ and $a=1$. As expected in a method that respects shift invariance and numerator-denominator consistency, the vector Ward identity~\refeq{vector} is satisfied.
The very same result is recovered if we directly write
\beq
\Pi_{\mu\nu}(p)= - \left[\int \frac{d^2 k}{4\pi^2} \, \text{tr} \left(\gamma_{\mu}\frac{1}{\slashed{k}-\slashed{p}+i\mu}\gamma_{\nu}\frac{1}{\slashed{k}+i\mu}\right) \right]^S,
\eeq
as the same $\mu^2$ term appears after the trace is evaluated.

Let us next compute the VA correlator:
\begin{align}
\Pi^5_{\mu\nu}(p) & =-\left[\int \frac{d^2 k}{4\pi^2} \, \text{tr} \left(\gamma_{\mu}\frac{1}{\slashed{k}-\slashed{p}}\gamma_{\nu}\gamma_{5}\frac{1}{\slashed{k}} \right) \right]^R \nn
& = - \left[\text{tr}\left(\gamma_{\mu}\gamma_\alpha\gamma_{\nu}\gamma_{5} \gamma_{\beta}\right) B_{\alpha\beta} \right]^R.
\end{align}
To evaluate the trace without ambiguities, we simply use the definition of $\gamma_5$~\refeq{gamma5def}. Then, refraining from using~\refeq{Schouten}, we have
\beq
\text{tr}\left(\gamma_{\mu}\gamma_\alpha\gamma_{\nu}\gamma_{5} \gamma_{\beta}\right)=
2\left(-\epsilon_{\beta\nu}\delta_{\alpha\mu}+
\epsilon_{\mu\nu}\delta_{\alpha\beta}-\epsilon_{\alpha\nu}\delta_{\beta\mu}+\epsilon_{\beta\alpha}\delta_{\mu\nu}
-\epsilon_{\mu\alpha}\delta_{\beta\nu}-\epsilon_{\beta\mu}\delta_{\alpha\nu}\right), \label{tracefour5}
\eeq
from which the normal form is readily obtained. 
Note that only the second term on the right-hand side of~\refeq{tracefour5} gives rise to $B_{\alpha\alpha}$, with contracted indices.  Using~\refeq{Bmunu} and~\refeq{Balal}, we get
\beq
\Pi^5_{\mu\nu}(p) = \frac{1}{\pi} \epsilon_{\nu\alpha} \left(\frac{p_\mu p_\alpha}{p^2} - \delta_{\mu\alpha} \right), 
\eeq
which agrees with~\refeq{simple5}. The vector Ward identity~\refeq{vectoraxial} and the anomalous axial one~\refeq{qaxial}, with $X=1/\pi$, follow. Observe that a different result, with the anomaly in the $\mu$ index, would have been obtained had we anticommuted the $\gamma_5$ with $1/\slashed{k}$. In fact, we can directly evaluate the left-hand side of~\refeq{qaxial}:
\begin{align}
p_\nu \Pi^5_{\mu\nu}(p) = - \left[\int \frac{d^2 k}{4\pi^2} \, \text{tr} \left(\gamma_{\mu}\frac{1}{\slashed{k}-\slashed{p}}(\slashed{p}- \slashed{k}+\slashed{k})\gamma_{5}\frac{1}{\slashed{k}} \right) \right]^R \nn
= 0 - 2 \left[\int \frac{d^2 k}{4\pi^2} \, \text{tr} \left(\gamma_{\mu}\frac{1}{\slashed{k}-\slashed{p}}
\hat{\slashed{k}} \gamma_5 \frac{1}{\slashed{k}} \right) \right]^R,
\end{align}
where the non-vanishing, evanescent term comes from the anticommutator $\{\slashed{k},\ga_5\}$, see~\refeq{anticonmutador}. Using the relation
\beq
\hat{\slashed{k}} \slashed{k} = k^2-\bar{k}^2 = \mu^2, \label{kkbar}
\eeq
an extra integral appears which gives the result~\refeq{qaxial}. 

Again, the same result can be obtained writing
\beq
\Pi^5_{\mu\nu}(p)  =- \left[\int \frac{d^2 k}{4\pi^2} \, \text{tr} \left(\gamma_{\mu}\frac{1}{\slashed{k}-\slashed{p}+i\mu}\gamma_{\nu} \gamma_5 \frac{1}{\slashed{k}+i\mu}\right) \right]^S . \label{quick5}
\eeq
As explained in the previous section, the presence of $\gamma_5$ should be obviated in assigning the relative signs of the $i\mu$ shifts. After writing~\refeq{quick5}, GnDI are allowed, and in particular we can anticommute $\gamma_5$ with the Dirac matrices. The origin of the anomaly can then be tracked to the extra integral arising from
\beq
\left\{ \gamma_5 , \slashed{k}-i \mu \right\} = - 2 i \mu \gamma_5 ,
\eeq
which is closely related to~\refeq{anticonmutador}.

Finally, let us calculate the AA correlator,
\begin{align}
\Pi^{55}_{\mu\nu}(p) & =-\left[\int \frac{d^2 k}{4\pi^2} \, \text{tr} \left(\gamma_{\mu} \gamma_5 \frac{1}{\slashed{k}-\slashed{p}}\gamma_{\nu}\gamma_{5}\frac{1}{\slashed{k}} \right) \right]^R \nn
& =- \left[\text{tr}\left(\gamma_{\mu} \gamma_5 \gamma_\alpha\gamma_{\nu}\gamma_{5} \gamma_{\beta} \right) B_{\alpha\beta} \right]^R.
\end{align}
First note that if we used $\gamma_5^{\text{AC}}$, we would immediately find $\Pi_{\mu\nu}^{55}=-\Pi_{\mu\nu}$, at odds with~\refeq{simple55}. But in our method we should not anticommute before the $\mu$ shift. The consistent result is obtained by using the definition~\refeq{gamma5def} for the two $\ga_5$. Then we need to evaluate a trace with eight Dirac matrices, contract with $B_{\alpha\beta}$ and use~\refeq{Bmunu} and~\refeq{Balal}. The computation is not difficult and gives the expected result,~\refeq{simple55}. A faster procedure is to make use of~\refeq{anticonmutador} and~$\ga_5^2=-1$ to write
\beq
\text{tr}\left( \ga_\mu \ga_5 (\slashed{k}-\slashed{p}) \ga_\nu \ga_5 \slashed{k} \right) = - \text{tr}\left( \ga_\mu (\slashed{k}-\slashed{p}) \ga_\nu \slashed{k} + 2 \ga_\mu \hat{\slashed{k}} \ga_\nu \kslash \right) .
\eeq
From this and~\refeq{kkbar} we easily obtain
\begin{align}
\Pi^{55}_{\mu\nu}(p) & = -\Pi_{\mu\nu} + 4 \delta_{\mu\nu} \left[\int \frac{d^2 k}{4\pi^2} \frac{\mu^2}{(k^2+\mu^2)^2}\right]^S \nn
& = -\Pi_{\mu\nu} -\frac{1}{\pi} \delta_{\mu\nu} \nn
& = -\frac{1}{\pi} \frac{p_\mu p_\alpha}{p^2} .
\end{align}
Once again, the same extra integral and therefore the same result are obtained by shifting the denominators with the prescribed signs,
\beq
\Pi^{55}_{\mu\nu}(p)  = - \left[\int \frac{d^2 k}{4\pi^2} \, \text{tr} \left(\gamma_{\mu} \gamma_5 \frac{1}{\slashed{k}-\slashed{p}+i \mu}\gamma_{\nu}\gamma_{5}\frac{1}{\slashed{k}+i\mu} \right) \right]^S .
\eeq
After this shift, which automatically performs the correct tensor identification, all the standard properties of $\gamma_5$ can be safely employed to simplify the calculation. Note that the very same procedure is followed in FDF.

The situation in $n=4$ is completely analogous, except for the fact that in that case the VA correlator studied here vanishes and the axial anomaly manifests itself in the familiar triangular diagrams.  These have been calculated in DReg~\cite{tHV}, consistent DRed~\cite{anomalyDRed}, CDR~\cite{AbelianCDR,CDRanomaly}, FDR~\cite{FDR}, CIReg~\cite{CIRanomaliesI} and~FDF~\cite{Gnendiger}. These calculations show that, as long as no GnDI is used before tensor identification, the vector Ward identities are automatically preserved and the anomaly is localized in the axial current. 

\subsection{Axial vertex Ward identity in four dimensions}
As an example with an open fermion chain, we consider the correlation function $\Gamma_\mu^5(p_1,p_2) = \langle j^5_\mu (p_1+p_2)  \bar{\psi}(-p_1) \psi(-p_2)  \rangle_{\text{1PI}}$ (with the Legendre transform applied only to the elementary fields) in four-dimensional\footnote{The corresponding diagrams in $n=2$ are finite by power counting and have no ambiguities.} massless QED, that is,
\beq
\mathcal{L} = \frac{1}{4} F_{\mu\nu} F_{\mu\nu} + \frac{1}{2} \d_\mu A_\nu \d_\mu A_\nu + \bar{\psi} \slashed{D} \psi ,  \label{QED}  
\eeq
with $D_\mu = \d_\mu - i e A_\mu$. As manifest in~\refeq{QED}, we work in the Feynman gauge. 
There is no anomaly associated to this correlator, i.e. the theory can be renormalized in such a way that the Ward identity
\beq
(p_1+p_2)_\mu \Gamma^5_\mu(p_1,p_2) = e \left(\ga_5 \Sigma(p_1) -  \Sigma(p_2) \ga_5 \right)\label{WI}
\eeq
is satisfied, with $\Sigma(p)=\langle \bar{\psi}(p) \psi(-p) \rangle_{\text{1PI}}$. However, it is known that this identity is not satisfied in DReg with the tHV definition of $\gamma_5$~\cite{tHV}. The reason is that  the GnDI $\pslash \gamma_5 = (\kslash + \pslash) \gamma_5-\gamma_5 \kslash$, which is needed in the combinatorial proof, does not hold for a non-anticommuting $\gamma_5$. The Ward identity can be recovered by adding a finite gauge-invariant counterterm. This is a necessity if the axial symmetry is gauged.

It is clear that the Ward identity~\refeq{WI} will also be violated in the consistent versions of DRed and implicit methods that employ the $\gamma_5$ definition in~\refeq{gamma5def}. Let us check this explicitly by one-loop calculations. Again, we use FDR for definiteness, but exactly the same results are found in CDR, CIReg and also in consistent DRed and FDF in $\overline{\text{MS}}$. The results in DReg are quantitatively different in this case. $\Sigma$ and $\Gamma_\mu$ in the following are understood to be the one-loop contributions to the corresponding correlation functions.

The fermion self-energy is given at one loop by 
\beq
\Sigma(p) = -i e^2\int \frac{d^4k}{(2\pi)^4} \gamma_{\alpha}\frac{1}{\slashed{k}}\gamma_{\alpha}\frac{1}{(k-p)^2},
\eeq
It has no potential ambiguity of the kind we are discussing. The result in FDR is easily found to be
\beq
\Sigma(p) = i \frac{e^2}{(4\pi)^2} \, \pslash \left(2-\log \frac{p^2}{\mu^2} \right) .
\eeq
Let us now compute the axial vertex $\Gamma^5_\mu$, which at one loop is given by 
\begin{align}
\Gamma^5_{\mu}(p_1,p_2) & = -i e^3 \left[\int \frac{d^4 k}{(2\pi)^4} \gamma_{\alpha}\frac{1}{\slashed{k}-\slashed{p}_2}\gamma_{\mu}\gamma_{5}\frac{1}{\slashed{k}+\slashed{p}_1}\gamma_{\alpha}\frac{1}{k^{2}} \right]^R \nn
& = -i e^3 \left[ S_{\alpha \beta \mu 5 \delta \alpha} C_{\beta \kappa}(p_1,p_2) \right]^R,
\end{align}
with
\beq
C_{\alpha\beta}(p_1,p_2) = \int \frac{d^4 k}{(2\pi)^4}  \frac{(k-p_2)_\alpha (k+p_1)_\beta}{k^2 (k-p_2)^2 (k+p_1)^2} .
\eeq
Substituting $\gamma_5$ by its definition~\refeq{gamma5def}, 
\beq
S_{\alpha \beta \mu 5 \kappa \alpha} = \frac{1}{4!} \epsilon_{\nu\rho\sigma\tau} S_{\alpha \beta \mu \nu\rho\sigma\tau \kappa \alpha} . 
\eeq
Next decompose $S_{\alpha \beta \mu \nu\rho\sigma\tau \delta \alpha}$ as in~\refeq{complete}. Since the index $\alpha$ is contracted, there are contributions proportional to $[\mu_1 \ldots \mu_m]$ with $m=1,3,5,7$. As pointed out before,  these combinations can be factored out of $[.]^R$, so the ones with $m=5,7$ can be directly set to zero, as in four genuine dimensions. Then, we contract indices with the resulting metrics and use the CIReg results
\begin{align}
[C_{\alpha\beta}(p_1,p_2)]^R & = \frac{1}{(4\pi)^2}\Big\{\frac{\delta_{\alpha\beta}}{4}\left[3 - p_{2}^2 \xi_{0,1} - p_{1}^2 \xi_{1,0} -  \log{\frac{(p_{1}+p_{2})^2}{\mu^2}}\right]\nonumber\\
&+\left[\left(\xi_{0,2}-\xi_{0,1}\right)(p_{1})_{\alpha}(p_{1})_{\beta}-\xi_{1,1}(p_{1})_{\alpha}(p_{2})_{\beta}+(p_{1}\rightleftharpoons p_{2},\xi_{m,n}\rightleftharpoons \xi_{n,m})\right]\nonumber\\
&-(p_{1})_{\beta}(p_{2})_{\alpha}\left(\xi_{0,0}-\xi_{0,1}-\xi_{1,0}\right)\Big\},\\ 
[C_{\alpha\alpha}(p_1,p_2)]^R & = \frac{1}{(4\pi)^2}\left[2-\frac{(p_{1} + p_{2})^2}{2} \xi_{0,0} - \frac{1}{2}\log{\frac{p_{2}^2}{\mu^2}} - \frac{1}{2}\log{\frac{p_{1}^2}{\mu^2}}\right]  ,
\end{align}
which in this massless case (and also in the massive case in the mass-independent version of CIReg) exactly coincide with the FDR ones. The functions $\xi_{n,m}\equiv\xi_{n,m}(p_{2},p_{1})$ are defined in the appendix. Importantly, the last integral includes the shift $k^2\to k^2+\mu^2$ in the numerator.  The final result is
\begin{align}
\label{result}
\Gamma_{\mu}^5(p_1,p_2)&=-i \frac{e^3}{(4\pi)^2}\bigg[\gamma ^{\mu }\gamma_{5}\Big[3 - (p_{1} + p_{2})^2\xi_{0,0}+ p_{2}^2 \xi_{0,1} + p_{1}^2 \xi_{1,0}- \log{\frac{p_{1}^2}{\mu^2}}-\log{\frac{p_{2}^2}{\mu^2}} \nonumber \\
& + \log{\frac{(p_{1}+p_{2})^2}{\mu^2}})\Big]
+2\Big\{\slashed{p_{2}}\gamma_{5} \big[p_{1}^{\mu } (2\xi_{1,1}- \xi_{0,1}-\xi_{1,0}- \xi_{0,0})+2p_{2}^{\mu } (\xi_{0,1}-\xi_{0,2})\big] \nn
& +(p_{1}\rightleftharpoons p_{2},\xi_{m,n}\rightleftharpoons \xi_{n,m})\Big\}
-2\left(\xi_{0,0}+\xi_{0,1}+\xi_{1,0}\right) \epsilon_{\delta\mu\alpha\beta}p_{2}^{\alpha}p_{1}^{\beta}\gamma^{\delta}\bigg] .
\end{align}
An equivalent procedure that simplifies the Dirac algebra is to anticommute the $\gamma_5$ to the right, using~\refeq{anticonmutador}. This leads to
\beq
\Gamma^5_{\mu}(p_1,p_2)  = - ie^3 \left [ (2\bar{\delta}_{\rho\kappa} - \delta_{\rho\kappa}) S_{\alpha \beta\mu \rho \alpha} C_{\beta \kappa}(p_1,p_2) \right]^R .
\eeq
Decomposing $S_{\alpha \beta\mu \rho \alpha}$ and using the rules~\refeq{tdelta} and~\refeq{bardeltaR}, we find again~\refeq{result}. 
Even more easily, the same result can be found fixing the $\mu$ terms from the very beginning with the same rule used above,
 \beq
 \Gamma^5_{\mu} (p_1,p_2) = -i e^3 \left[\int \frac{d^4 k}{(2\pi)^4} \gamma_{\alpha}\frac{1}{\slashed{k}-\slashed{p}_2+i\mu}\gamma_{\mu}\gamma_{5}\frac{1}{\slashed{k}+\slashed{p}_1+i\mu}\gamma_{\alpha}\frac{1}{k^{2}} \right]^S  \label{directvertex}
 \eeq
After this, $\gamma_5$ can be safely anticommuted with the Dirac matrices (and commuted with $\mu$). Let us note again that this same prescription is used in FDF, so the result will be identical in that method. Even if the last procedure looks simpler, it should be noted that it is less universal than the other ones, as we have pointed out in the previous section.

The result~\refeq{result} does not satisfy the Ward identity~\refeq{WI}. Instead, using the relations in the appendix we find
\beq
(p_{1}+p_{2})_{\mu}\Gamma_{\mu}(p_1,p_2)= e \left( \gamma_5 \Sigma(p_1)-  \Sigma(p_2) \gamma_5  \right) - 2 i \frac{e^3}{(4\pi)^2} (\pslash_1+\pslash_2)\gamma_5 . \label{nonWI}
\eeq
To isolate the origin of the extra local term, we can compute the left-hand side of~\refeq{nonWI} directly. For instance, using the expression in~\refeq{directvertex},
\begin{align}
(p_{1}+& p_{2})_{\mu}  \Gamma_{\mu}(p_1,p_2) & \nn
& =  -i e^3 \left[\int \frac{d^4 k}{(2\pi)^4} \gamma_{\alpha}\frac{1}{\slashed{k}-\pslash_2+i\mu}(\pslash_1+\kslash +i \mu + \pslash_2 - \kslash-i \mu) \gamma_{5}\frac{1}{\slashed{k}+\pslash_1+i\mu}\gamma_{\alpha}\frac{1}{k^{2}} \right]^S \nn 
& = e \left(\gamma_5 \Sigma(p_1) -  \Sigma(p_2) \gamma_5 \right) - i e^3 \left[\int \frac{d^4 k}{(2\pi)^4} \gamma_{\alpha}\frac{1}{\slashed{k}-\pslash_2+i\mu}(2i \mu \gamma_5) \frac{1}{\slashed{k}+\pslash_1+i\mu}\gamma_{\alpha} \frac{1}{k^{2}} \right]^S .
\end{align}
It can be checked that the extra integral above gives the extra local term on the right-hand side of~\refeq{nonWI}. The axial symmetry can be restored by canceling this term with a finite counterterm proportional to $\bar{\psi} \slashed{B} \gamma_5 \psi$, where $B_\mu$ is a source coupled to $j_\mu^5$.

Our results are consistent with the ones in~\cite{Gnendiger}, where $\Gamma_{\mu}$ is calculated for massive on-shell fermions in FDH with a tHV $\gamma_5$ and FDF, which give the same result, and in FDH with $\gamma_5^{\text{AC}}$, which differs by a local term.  In the context of dimensional methods, it has been observed that identity~\refeq{WI} and similar Ward identities can be preserved by moving all $\gamma_5$ to one end of open fermion lines before regularization and renormalization~\cite{Tsai}. The reason is that, by doing this, the $\gamma_5$ does not interfere with the necessary identity in the combinatorial proof. This is not quite the same as using $\gamma_5^{\text{AC}}$, as the $\gamma_5$ matrices are not allowed to be anticommuted to an arbitrary position. This trick works equally well in implicit methods and it has actually been advocated in FDR~\cite{FDR}. Observe, nevertheless, that this procedure goes beyond the basic idea in these methods of substituting the bare expressions, in the form obtained from the Feynman rules, by their renormalized value. A previous non-trivial manipulation is performed. Then, one needs to check that this does not interfere with unitarity or with the quantum action principle in multiloop calculations.

\section{Conclusions}
\label{s:conclusions}
In the last decade, there has been a renewed interest in alternative methods to perform perturbative calculations in quantum field theory (see \cite{review} for a recent review). This has been motivated by the increasing complexity of the computations required to match the sensitivity of present and future experiments and by the development of new techniques for on-shell scattering amplitudes,  based on unitarity and analyticity.  The most efficient methods are either variations of dimensional regularization or implicit methods in fixed dimension, which act directly on the bare integrals, often at the integrand level, and do not need to keep track of counterterms. Besides other possible advantages, the latter could be expected to handle more easily chiral theories, such as the Standard Model, since the concept of chirality is dimension specific. In this paper we have examined this issue in implicit fixed-dimension methods such as CDR, CIReg and FDR. We have shown that, somewhat counterintuitively, the difficulties one has to address in these methods are very similar to the ones in dimensional methods. They can be dealt with in a similar manner.

The origin of these difficulties is the fact that contraction of Lorentz indices does not commute with renormalization in these implicit methods. We have observed that this is actually required to preserve both shift invariance and numerator-denominator consistency, which are the crucial ingredients in the perturbative proof of the quantum action principle. The latter leads to Ward identities of local and global symmetries in the renormalized theory. But it turns out that this non-commutation property is incompatible with certain identities, specific to the ordinary $n$-dimensional space in which a given theory is defined. Hence, a na\"ive usage of these identities may lead to inconsistencies. The situation is similar to the one in dimensional methods. And a way out is also to simply avoid using these identities before renormalization. This statement can be made more formal by defining the theory in an infinite dimensional space QnS, which only shares a few features with the real $n$-dimensional space.

Working in QnS is necessary for consistency, but it brings about some complications in theories with fermions. First, it turns out that the standard Dirac algebra cannot be preserved in odd dimensions. Possible solutions to this problem will be investigated elsewhere. Second, there is no finite complete set in Dirac space and hence the standard Fierz identities do not hold. One consequence of this is that these methods break supersymmetry. And third, we have argued that it is impossible to define a unique $\gamma_5$ matrix that anticommutes with the Dirac matrices and reduces to the standard $\gamma_5$ after renormalization (or in finite expressions). We have then proposed to use the standard explicit definition with the antisymmetric $\epsilon$ tensor in terms of the Dirac matrices. This is similar to the t'Hooft-Veltman definition in dimensional regularization and has the same consequences. Axial anomalies are reproduced, but in addition some spurious anomalies emerge, which should be removed {\it a posteriori\/} by local counterterms, added by hand. This is equivalent to the direct use of an anticommuting $\gamma_5$, when allowed~\cite{Jegerlehner:2000dz}. 

In the implicit methods, it is also necessary to discriminate between different tensor structures. To avoid ambiguities in this discrimination, we have proposed a systematic renormalization procedure, in which the expressions to be renormalized are first  put in a certain normal form, using only relations valid in QnS. We have also suggested some simplifications that make use of the decomposition $\text{QnS=GnS}\oplus X$, where GnS is the genuine $n$-dimensional space. The advantage of introducing this direct-sum structure is that it allows to use standard identities in GnS at some steps of the calculations.

We have mostly studied renormalization of implicit fixed-dimension methods at the one-loop level and have only made some suggestions about how our consistent procedure should be extended to higher orders. Our suggestions seem related to the requirement of sub-integration consistency in FDR. A more systematic analysis of this, and more generally of renormalization to all orders, would be very interesting.

In the context of chiral theories, we have also reconsidered shortcuts that exist at one loop and in simple higher-loop diagrams in FDR, which allow to discriminate the tensor structures from the very beginning and obtain the same results in a more direct way. A generalization of these shortcuts to arbitrary diagrams would allow to reduce computational cost of heavy calculations. We think that the ideas in FDF can be helpful in this regard.

Finally, we have treated chiral theories in a formalism with Dirac spinors and chiral projectors. It would be interesting to see how our findings are translated to calculations with Weyl spinors and in superspace~\cite{Dreiner:2008tw}.

\section*{Acknowledgments}
We thank Alice Donati, Ben Page and Roberto Pittau for sharing their expertise in FDR, and Paco del Aguila for a critical reading of the manuscript. M.P.V.\ also thanks Martinus Veltman for discussions of related topics in dimensional regularization, which have partly motivated this work.
The work of A.M.B.\  and M.P.V.\ has been supported by the Spanish MINECO project FPA2016-78220-C3-1-P (Fondos FEDER) and the Junta de Andaluc\'{\i}a grant FQM101. The work of M.P.V.\ has also been supported by the European Commission, through the contract PITN-GA-2012-316704 (HIGGSTOOLS).
A.L.C.\ acknowledges financial support from CAPES (Coordena\c{c}\~{a}o de Aperfei\c{c}oamento de
Pessoal de N\'{i}vel Superior), Brazil.

\section*{Appendix} \label{A}

In this appendix, we collect the finite three-points functions used in the evaluation of the axial vertex in four dimensions. 
We define the functions $\xi_{nm}\equiv\xi_{nm}(p_{2},p_{1})$ as
\begin{equation}
\xi_{nm}(p_{2},p_{1})=\int^1_0 dz\int^{1-z}_0 dy \frac{z^n y^m}{Q(y,z)},\\
\end{equation}
with
\begin{equation}
Q(y,z)=[\mu^2 - p_{2}^2 y(1-y) -p_{1}^2 z(1-z) - 2(p_{2}\cdot p_{1})yz],
\end{equation}
and notice that these functions have the property $\xi_{nm}(p_{2},p_{1})=\xi_{mn}(p_{1},p_{2})$.
Using integration by parts \cite{Orimar}, the relations below follow
\begin{align}
&p_{1}^2 \xi_{11}-(p_{2}\cdot p_{1})\xi_{02}=\frac{1}{2}\left[ -\frac{1}{2}\log{\frac{(p_{1}+p_{2})^2}{\mu^2}}+\frac{1}{2}\log{\frac{p_{2}^2}{\mu^2}}+p_{1}^2 \xi_{01}
\right],\\
&p_{2}^2 \xi_{11}-(p_{2}\cdot p_{1})\xi_{20}=\frac{1}{2}\left[ -\frac{1}{2}\log{\frac{(p_{1}+p_{2})^2}{\mu^2}}+\frac{1}{2}\log{\frac{p_{1}^2}{\mu^2}}+p_{2}^2 \xi_{10}
\right],\\
&p_{1}^2 \xi_{10}-(p_{2}\cdot p_{1})\xi_{01}=\frac{1}{2}\left[ -\log{\frac{(p_{1}+p_{2})^2}{\mu^2}}+\log{\frac{p_{2}^2}{\mu^2}}+p_{1}^2 \xi_{00}\right],\\
&p_{2}^2 \xi_{01}-(p_{2}\cdot p_{1})\xi_{10}=\frac{1}{2}\left[ -\log{\frac{(p_{1}+p_{2})^2}{\mu^2}}+\log{\frac{p_{1}^2}{\mu^2}}+p_{2}^2 \xi_{00}\right],\\
&p_{1}^2 \xi_{20}-(p_{2}\cdot p_{1})\xi_{11}=\frac{1}{4}\left[-1 + p_{2}^2\xi_{01} + 3p_{1}^2\xi_{10}\right],\\
&p_{2}^2 \xi_{02}-(p_{2}\cdot p_{1})\xi_{11}=\frac{1}{4}\left[-1 + p_{1}^2\xi_{10} + 3p_{2}^2\xi_{01}\right].
\end{align}

\end{document}